\documentclass{iau}
\usepackage{graphicx,natbib,amsmath}
 
\newcommand{\apj}{ApJ}           
\newcommand{\apjl}{ApJ}           
\newcommand{\mnras}{MNRAS}       
\newcommand{\nat}{Nature}
\newcommand{\aap}{A\&A}
\newcommand{\araa}{ARA\&A}
\newcommand{\aj}{AJ}
\newcommand{\pasp}{PASP}
\newcommand{\apjs}{ApJS}           

\newcommand{\atl}{ATLAS$^{\rm 3D}$}
\newcommand{\kms}{\hbox{km s$^{-1}$}}
\newcommand{\msun}{\hbox{$M_\odot$}}
\newcommand{\re}{\hbox{$R_{\rm e}$}}

\newcommand{\reffig}[1]{Fig.~\ref{#1}}
\newcommand{\refeq}[1]{equation~(\ref{#1})}

\title[Dynamical masses of early-type galaxies at $z\sim2$]
{Dynamical masses of\\ early-type galaxies at $z\sim2$}

\author{Michele Cappellari}

\affiliation{Sub-department of Astrophysics, Department of Physics, University of Oxford\\ Denys Wilkinson Building, Keble Road, Oxford OX1 3RH \\ email: {\tt cappellari@astro.ox.ac.uk}}

\pubyear{2013}
\volume{295}
\jname{The intriguing life of massive galaxies}
\editors{D. Thomas, A. Pasquali \& I. Ferreras, eds.}
\begin{document}

\maketitle

\begin{abstract}
The evolution of masses and sizes of passive (early-type) galaxies with redshift provides ideal constraints to galaxy formation models. These parameters can in principle be obtained for large galaxy samples from multi-band photometry alone. However the accuracy of photometric masses is limited by the non-universality of the IMF. Galaxy sizes can be biased at high redshift due to the inferior quality of the imaging data. Both problems can be avoided using galaxy dynamics, and in particular by measuring the galaxies stellar velocity dispersion. Here we provide an overview of the efforts in this direction.
\keywords{galaxies: elliptical and lenticular, cD –- galaxies: evolution -– galaxies: formation -– galaxies: kinematics and dynamics –- galaxies: structure.}
\end{abstract}

\firstsection
\section{Introduction}

In the era of precision cosmology we think we can accurately predict the distribution of dark matter in the Universe \citep{Springel2005nat}. Dark matter is thought to be the main driver for the hierarchical assembly of galaxy stellar masses. However the mechanism by which the gas sinks into the dark matter potential wells and form stars is still largely unknown, due to the complex and difficult to model effect of baryonic physics. For this reason our understanding of galaxy formation must be driven by observations.

\section{Uncertainties of galaxy stellar masses}

A first key observable to compare with galaxy formation models is the galaxy stellar mass, which is expected to monotonically growth with time during galaxy mergers. In recent years confidence started to growth about our ability to measure stellar masses with a few tens of percent accuracy \citep[e.g.][]{Gallazzi2009,Maraston2010}, up to redshift $z\sim2$ and beyond, thanks to the availability of multi-bands photometric surveys with the Hubble Space Telescope (HTS) \citep[e.g.][]{Giavalisco2004} combined with detailed stellar population models \citep[e.g.][]{bruzual03,Maraston2005,Vazdekis2010,Conroy2012models}. 

Mass determinations using stellar population models depend on the assumption of the stellar initial mass function (IMF). Until recently it appeared sensible to assume the IMF is universal in different galaxies, and of \citet{Kroupa2001} or \citet{Chabrier2003} type. This assumption was motivated by the observed IMF universality in different environments within our own Galaxy \citep{Kroupa2002,Bastian2010}. 

Recent results from spectral absorption features \citep[e.g.][]{vanDokkum2010} and gravitational lensing \citep[e.g.][]{Auger2010imf} found indications for a heavier IMF in massive early-type galaxies (ETGs), more consistent with an IMF with the \citet{Salpeter1955} logarithmic slope or even heavier, placing the IMF universality into question. The extensive dynamical modelling effort of \citet{Cappellari2012} discovered a systematic trend in the IMF of ETGs. They found a mass normalization for the IMF varying between values consistent with a Kroupa/Chabrier IMF to heavier than a Salpeter IMF (assuming the population models are correct) as a function of either the mass-to-light ratio ($M/L$) or the galaxy stellar velocity dispersion $\sigma$. This trend implies potential systematic errors of up to a factor 2--3 in the galaxies stellar masses derived using stellar population models, even from optimal data.

\section{Uncertainties of galaxy sizes}

A second observable that is often used to constrain galaxy formation scenarios, in combination with galaxy mass, is the galaxy size, commonly parametrized by the projected half-light radius \re. The evolution of galaxy size with its mass depends sensitively on the galaxy assembly mechanism (see \citealt{Ciotti2009} for a comprehensive review). \re\ increases nearly proportionally to mass in the case of a galaxy assembly process via gas-poor major (1:1) mergers \citep{Hernquist1993}, due to energy conservation, while \re\ increases more rapidly during gas-poor minor (1:3 or less) merging \citep{Nipoti2003,Boylan-Kolchin2006}. \re\ {\em decreases} during gas-rich mergers or cold accretion, when gas accumulates towards the centre \citep[e.g.][]{Mihos1994burst}.

Massive (stellar mass $M_\star\gtrsim10^{11}$ M$_\odot$) passive galaxies at $z\sim2$ were found to have much smaller \re\ than their local counterparts of the same mass \citep{Daddi2005,diSeregoAlighieri2005,Trujillo2006,Longhetti2007,Toft2007,Cimatti2008,vanDokkum2008}. This suggested a scenario in which the progenitors of today's massive ETGs assemble most of their mass via minor dry mergers \citep[e.g.][]{Naab2009,Bezanson2009,Hopkins2009compact}, to account for the rapid size increase.

Unfortunately the half-light radius \re\ is an observationally ill-defined quantity. By definition measuring \re\ requires an extrapolation of the galaxy surface brightness profile to infinite radii, to estimate the galaxy {\em total} stellar light. This seemed a sensible approach a few decades ago, when the technique was used to quantify the sizes of elliptical galaxies, which were thought to be all well described by homologous $R^{1/4}$  \citet{deVaucouleurs1948} surface-brightness profiles. But we now know that ETGs are best represented by \citet{Sersic1968} profiles, with concentration increasing with galaxy mass \citep{Caon1993}. Moreover the \atl\ volume-limited ($M_\star\gtrsim6\times10^9$ M$_\odot$) survey of nearby ETGs \citep{Cappellari2011a} has revealed that the majority of the systems are more closely related to spiral galaxies than to genuine spheroidal ellipticals \citep{Cappellari2011b}. In fact as much as 2/3 of the galaxies classified as ellipticals are dominated by stellar disks, which are often missed by the photometry, but are clearly revealed by their integral-field stellar kinematics \citep{Krajnovic2011,Emsellem2011}. 

\begin{figure}
\centering
\includegraphics[width=\textwidth]{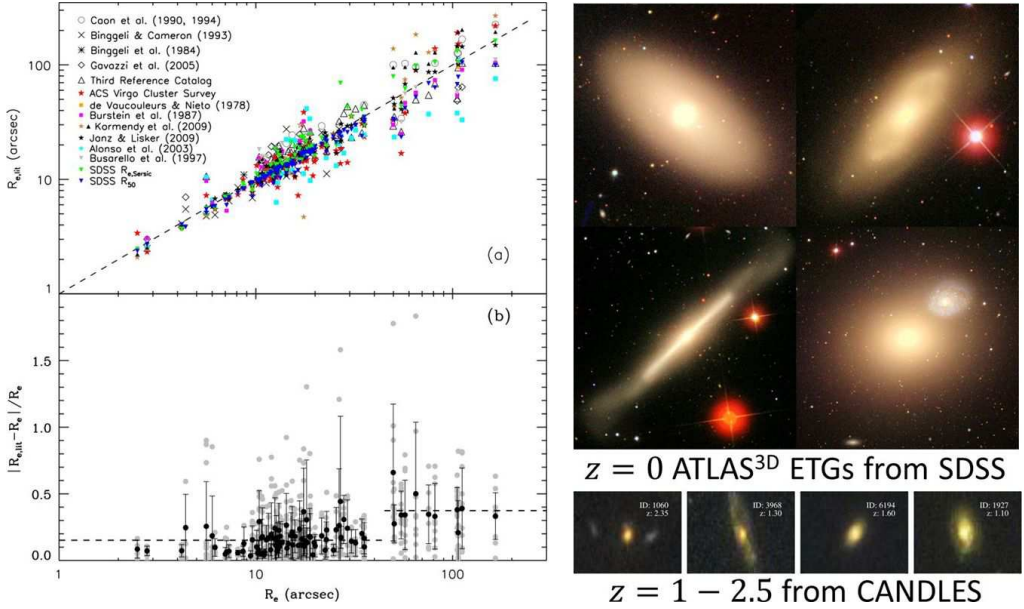} 
\caption{{\bf Left: Uncertainty of \re\ for galaxies in the Virgo cluster.} {\em Top Panel:} Different symbols compare the SDSS half-light radii from \citet{Chen2010} with independent determinations of the same quantity from the literature. {\em Bottom Panel:} Gray dots show the fractional error in the effective radius measurements, plotted as a function of the SDSS effective radii; the mean and standard deviation for individual galaxies are plotted as the black dots with error bars. Differences in \re\ of a factor $2\times$ are common, especially for the largest galaxies. (taken from \citealt{Chen2010}).
{\bf Right: Comparing images of nearby and distant ETGs.} {\em Top Panels:} true colour images of nearby ETGs from the \atl\ sample as imaged by the SDSS survey \citep{Abazajian2009}. {\em Bottom Panels:} true colour images of galaxies at $z=1-2.5$ from the CANDLES survey (taken from \citealt{Szomoru2012})}
   \label{fig:chen2010_fig11}
\end{figure}

The complex and multi-component nature of ETGs makes it impossible to know how the outer unobservable surface brightness profile should be extrapolated. Different authors use photometry of different quality and make different choices for the profile extrapolation, as well as for its extraction. For these reason, even from very good quality photometry of nearby galaxies, \re\ is often not known with an accuracy better than about a factor $2\times$, with errors dominated by systematics effects. This is illustrated in the left panel of \reffig{fig:chen2010_fig11} (from \citealt{Chen2010}), which provides a comparison between independent determinations of \re\ for well-studied ETGs in the Virgo cluster. 

Uncertainties and systematic biases in \re\ are likely to be more severe when galaxies at high redshift are compared to local ones. Differences in the measurement approach combine with (i) differences in the restframe wavelength of the observations, (ii) variations in the depth of the photometry (e.g.\ due to cosmological surface-brightness dimming) and (iii) inferior spatial sampling of the imaging of distant galaxies (see the right panel of \reffig{fig:chen2010_fig11}), (iv) possible colour gradients and nuclear AGN activity, which are generally stronger in the early Universe.

\section{Local benchmark for dynamical scaling relations}

\begin{figure}
\centering
 \includegraphics[width=0.8\textwidth]{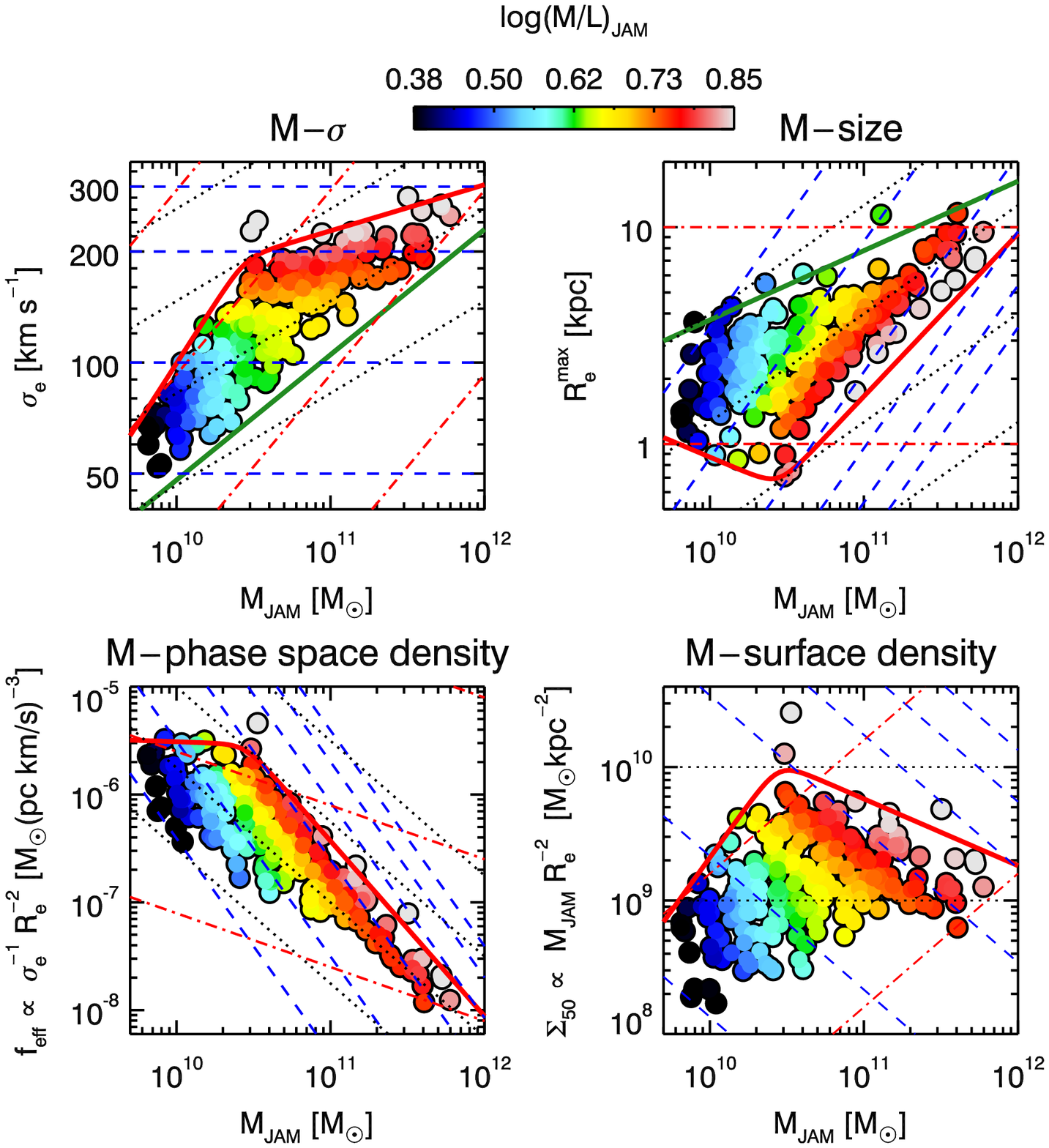} 
 \caption{{\bf The Mass Plane and it projections.} The top two panels show the $(M_{\rm JAM},\sigma_e)$ and $(M_{\rm JAM},R_e^{\rm max})$ coordinates. Overlaid are lines of constant $\sigma_e=50,100,200,300,400,500$ \kms\ (dashed blue), constant $R_e^{\rm max}=0.1,1,10,100$ kpc (dot-dashed red) and constant $\Sigma_e=10^8,10^9,10^{10},10^{11}$ \msun\ kpc$^{-2}$ (dotted black) {\em predicted} by the virial \refeq{eq:virial}. The observed $(M_{\rm JAM},\sigma_e,R_e^{\rm max})$ points follow the relation so closely that the coordinates provide a unique mapping on these diagram and one can reliably infer characteristics of the galaxies from any individual projection. In each panel the galaxies are coloured according to the (LOESS smoothed) $\log (M/L)_{\rm JAM}$ values, as shown in the colour bar. Moreover in all panels the thick red line shows the same ZOE relation projected via \refeq{eq:virial}. The green line is the $M-\re$ relation for late spiral galaxies (equation~3 from \citealt{Cappellari2011a}), which approximately defines the boundary where ETGs disappear. (Taken from \citealt{Cappellari2012p20})}
   \label{fig:cappellari2012_fig1}
\end{figure}

An alternative to measuring galaxy sizes to study the evolution of galaxy densities consists of extracting the galaxies velocity dispersion $\sigma$. This is more difficult to obtain than sizes, as it requires spectroscopic data (resolution $R\approx1000$) rather than imaging alone. However $\sigma$ does not suffer from the biases of \re\ and is closely related to it via the scalar virial equation \citep[e.g.][]{Cappellari2006}
\begin{equation}
\sigma^2 \approx \frac{G\, M_{\rm dyn}}{5.0\,\re}.
\label{eq:virial}
\end{equation}
Here $M_{\rm dyn}=2\times M_{1/2}\approx M_\star$, where $M_{1/2}$ is the mass within a sphere enclosing half of the total galaxy light, and the last approximation is due to the fact that $M_{1/2}$ is dominated by the stellar mass \citep{Cappellari2012p19}. The factor 5.0 was calibrated for \re\ measured in the classic way \citep{Burstein1987,deVaucouleurs1991,Jorgensen1995phot} using fixed \citet{deVaucouleurs1948} $R^{1/4}$ growth curves to extrapolate the outer profiles and using typical photometry of nearby galaxies.

To study the evolution of galaxy parameters with redshift, a reliable local benchmark is essential. In \reffig{fig:cappellari2012_fig1} we show the $(M,\sigma)$ and $(M,\re$) (and other projections) of the mass plane $(M,\re,\sigma)$ of ETGs obtained by the \atl\ project via detailed dynamical modelling of 260 galaxies \citep{Cappellari2012p20}. The study found that, when accurate and unbiased masses are used, ETGs satisfy \refeq{eq:virial} quite accurately. Different two-dimensional projections of the plane contain the same amount of information except for a coordinate transformation. This confirms in particular that one can reliably estimate $\sigma$ from the knowledge of $M$ and \re\ using \refeq{eq:virial}. The study also found that the zone of exclusion (ZOE) defined by local ETGs in the $(M,\re)$ plane shows a clear break at a characteristic mass $M_{\rm dyn}\approx3\times10^{10}$ M$_\odot$, with a corresponding break in the ZOE in the $(M,\sigma)$ plane and other projections. 

Indications for a change of slope in the \citet{Faber1976} relation between galaxy luminosity and $\sigma$ \citep{Davies1983} or in the \citet{Kormendy1977} relation between luminosity and surface brightness \citep{Binggeli1984} were found some time ago. However this is the first time the break is presented using dynamical masses instead of luminosities, and it is shown to accurately represent two projections of the same ZOE on the galaxies mass plane. This break is interpreted as evidence for a change in the galaxy accretion mechanism between bulge growth (in-situ star formation) and gas-poor merging (external accretion).

\section{Importance of $\sigma$ determinations at high-z}

\begin{figure}
\centering
 \includegraphics[width=0.8\textwidth]{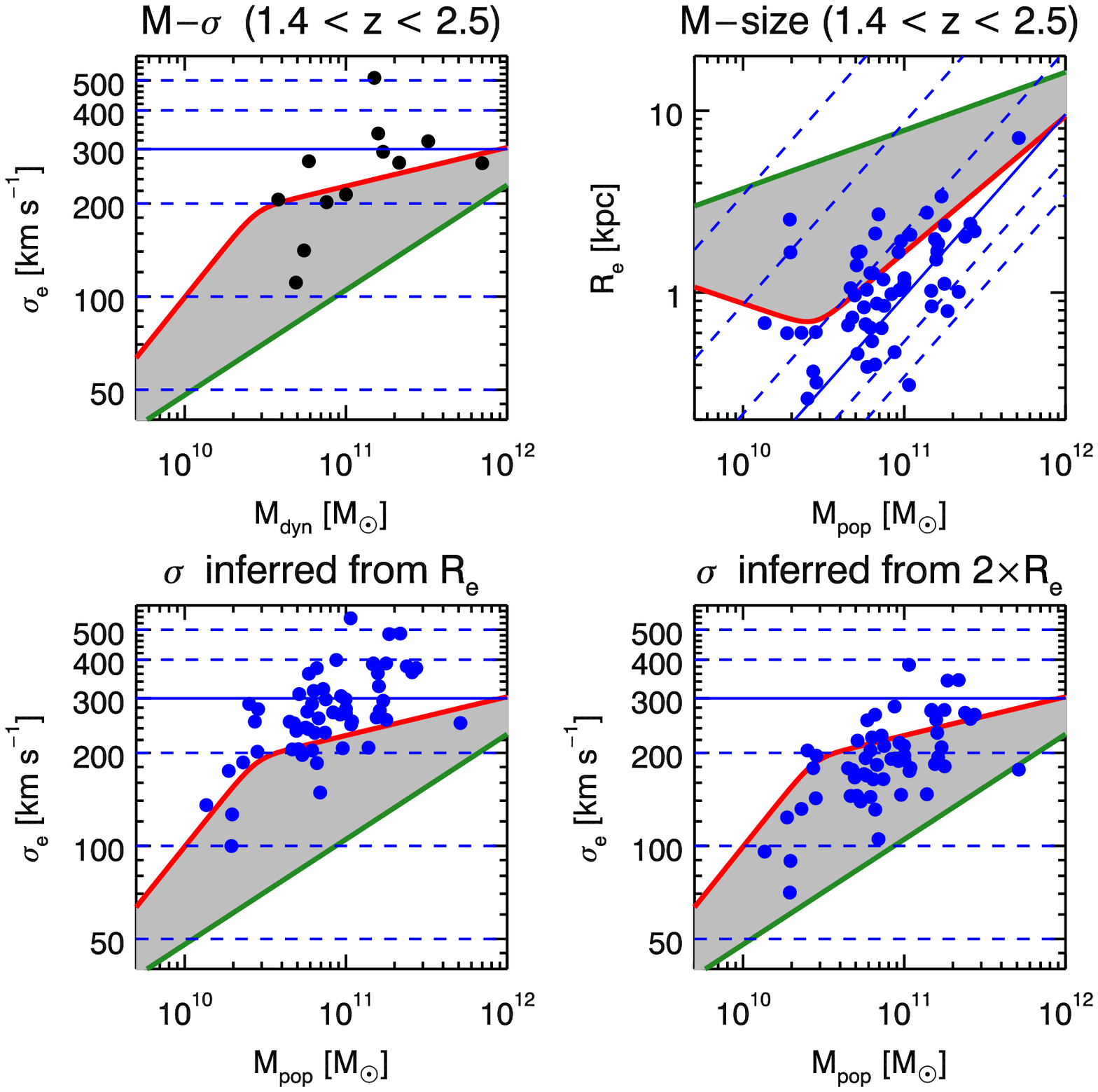} 
 \caption{{\bf Mass, size and $\sigma$ distributions at $z\sim2$.} {\em Top Left:} The observed $(M,\sigma)$ determinations (black filled circles) for ETGs with $1.4<z<2.5$ (from  \citealt{Cenarro2009,Cappellari2009,vanDokkum2009nat,vandeSande2011,Newman2010,Onodera2012,Toft2012}) is overlaid to the locus of local ETGs from \atl\ (grey region). The red and green thick lines are the same as in \reffig{fig:cappellari2012_fig1}. {\em Top Right:} as in the previous panel, for the distribution of $(M,\re)$  determinations (blue filled circles) (from \citealt{Cassata2011,Szomoru2012}). {\em Bottom Left:} The $(M,\sigma)$ distribution {\em predicted} using \refeq{eq:virial} from the values in the top-right panel. {\em Bottom Right:} as in the previous panel, but using $2\re$ to estimate $\sigma$ instead of the observed \re.}
   \label{fig:mass_sigma_size_z2}
\end{figure}

In the top-left panel of \reffig{fig:mass_sigma_size_z2} we show the available determinations of $\sigma$ for galaxies in the redshift interval $1.4<z<2.5$ \citep{Cenarro2009,Cappellari2009,vanDokkum2009nat,vandeSande2011,Newman2012,Onodera2012,Toft2012} and compare it with the location defined by nearby \atl\ ETGs. There is an indication for the high-redshift $\sigma$ determinations to lie generally above the locus of local galaxies with the same mass, as found earlier, and as expected for more compact systems. However, with the exception of one galaxy, the differences are relatively modest and are mostly consistent with local galaxies within the measurement errors.

In the top-right panel of \reffig{fig:mass_sigma_size_z2} we show state-of-the-art size determinations of \re\ for passive galaxies \citep{Cassata2011,Szomoru2012}, in the same redshift range. They were obtained from deep HST/WFC3 observations in the GOODS and CANDLES fields. The photometric samples are much larger than the one with $\sigma$ determinations, and in this case the difference between the locus of local ETGs and the $z\sim2$ ones is more dramatic, with the high-z galaxies having much smaller sizes than local ETGs.

In the bottom-left panel of \reffig{fig:mass_sigma_size_z2} the observed $(M,\re)$ distribution of the $z\sim2$ ETGs is converted into the $(M,\sigma)$ plane via \refeq{eq:virial}. This shows that the observed distribution of galaxy sizes implies more extreme $\sigma$ values than the ones that have been found so far. The expected $\sigma$ may even be larger than predicted here, due to the fact that high-z galaxies not only are observed to be smaller, but also tend to have steeper profiles than their local counterparts \citep[e.g.][]{Szomoru2012}. One way to bring the high-z predicted $\sigma$ values into better qualitative agreement with the few available determinations, would be to assume the \re\ values at $z\sim2$ are underestimated by about a factor of $2\times$ (bottom-right panel of \reffig{fig:mass_sigma_size_z2}).

However the comparison between the currently available $\sigma$ and \re\ determinations at $z\sim2$ is not sufficient to conclude that the high-z determinations of \re\ are underestimated. The sample of galaxies with available $\sigma$ is still too small to draw robust conclusions, and it is selected in a very different way than the more complete photometric samples. Moreover the galaxies with available $\sigma$ generally satisfy \refeq{eq:virial} reasonably well. What the comparison does emphasizes is the importance of obtaining more good-quality $\sigma$ values of high-z galaxies (see e.g.\ van de Sande in these proceedings), until a fully consistent picture emerges. 

The possibility of biases in the \re\ determinations of high-z galaxies also emphasizes the danger of using the local virial \refeq{eq:virial} for quantitative estimates of $M/L$, e.g.\ to study the evolution of the Fundamental Plane or to explore possible variations of the IMF with redshift. There is a clean solution to the problem. It consists of using detailed axisymmetric dynamical models of high-z galaxies \citep{vanderMarel2007,vanderWel2008vdM,Cappellari2009}, which describe in detail the shape and surface brightness profiles of the galaxies. Contrary to the dynamical $M/L$ inferred from the virial relation, the $M/L$ derived via dynamical models is virtually insensitive to even extreme changes in the surface brightness distribution outside the region where stellar kinematics is available. Although the correctness of this fact is best demonstrated with numerical experiments \citep[e.g.][]{Cappellari2009}, the reason for this robustness is similar to the one for which, in the spherical limit, the knowledge of the orbit of a single star is sufficient to infer the mass enclosed inside its orbit, irrespective of the large-scale mass distribution.

\begin{figure}
\centering
 \includegraphics[width=0.8\textwidth]{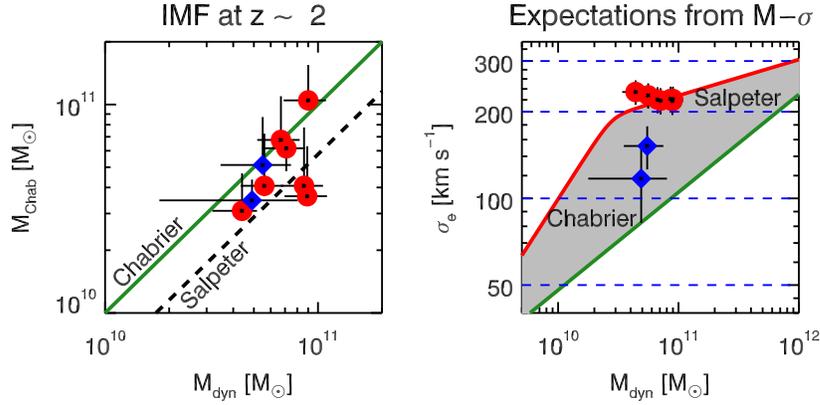} 
 \caption{{\bf Constraining the IMF at $z\sim2$.} {\em Left Panel:} dynamical versus population masses determined using a Chabrier IMF. Blue diamonds are galaxies with individually measured $\sigma$, while red filled circles are galaxies for which $\sigma$ from a stacked spectrum was assumed. The green solid line is the one-to-one relation. The dashed line is the one-to-one relation if the IMF was of Salpeter type. The latter is inconsistent with some of the determinations, indicating that some of the object require a Chabrier-like IMF (taken from \citealt{Cappellari2009}). {\em Right Panel:} Location of the galaxies in the left panel in the $(M,\sigma)$ plane. The grey area and thick red and green lines are as in \reffig{fig:cappellari2012_fig1}. The labels show the IMF normalization expected at different locations on this plane for local ETGs \citep{Cappellari2012p20}.}
   \label{fig:imf_z2}
\end{figure}

A demonstration of this modelling approach is given in the left panel of \reffig{fig:imf_z2}, which shows a first attempt at measuring the IMF normalization for galaxies at $z\sim2$ \citep{Cappellari2009}. The study used dynamical models to reproduce $\sigma$ determinations and concluded that some of the high-z ETGs are {\em required} to have a mass normalization of the IMF as low as Kroupa/Chabrier, like local ETGs. However the study allows for the possibility of some of the studies galaxies to have a heavier Salpeter-like mass normalization. This is not inconsistent with local findings, which indicate the IMF should vary between Kroupa to Salpeter in the interval $\sigma\approx80-260$ \kms (right panel of \reffig{fig:imf_z2}; \citealt{Cappellari2012p20}). But the current data and samples are still not sufficient to draw strong conclusions on the IMF variation at high-z.

The near future looks bright for dynamical studies of galaxies at $z\sim2$, thanks to the availability of multi-object Spectrometers like MOIRCS on the Subaru telescope, the upcoming MOSFIRE on the Keck telescope and KMOS on the Very Large Telescope. These are optimized for multiplexed near-infrared observations of distant galaxies and should provide in a few years numerous $\sigma$ determinations at high redshift ETGs.

\vspace{-0.5cm}

\section*{Acknowledgements}

\noindent
I acknowledge support from a Royal Society University Research Fellowship.


\begin{thebibliography}{62}
\expandafter\ifx\csname natexlab\endcsname\relax\def\natexlab#1{#1}\fi

\bibitem[{{Abazajian} {et~al}\mbox{.}(2009){Abazajian}, {Adelman-McCarthy},
  {Ag{\"u}eros}, {Allam}, {Allende Prieto}, {An}, {Anderson}, {Anderson},
  {Annis}, {Bahcall}, \& et~al.}]{Abazajian2009}
{Abazajian} K.~N. {et~al.}, 2009, \apjs, 182, 543

\bibitem[{{Auger} {et~al}\mbox{.}(2010){Auger}, {Treu}, {Gavazzi}, {Bolton},
  {Koopmans}, \& {Marshall}}]{Auger2010imf}
{Auger} M.~W., {Treu} T., {Gavazzi} R., {Bolton} A.~S., {Koopmans} L.~V.~E.,
  {Marshall} P.~J., 2010, \apjl, 721, L163

\bibitem[{{Bastian} {et~al}\mbox{.}(2010){Bastian}, R., \&
  {Meyer}}]{Bastian2010}
{Bastian} N., R. C.~K., {Meyer} M.~R., 2010, \araa, 48, 339

\bibitem[{{Bezanson} {et~al}\mbox{.}(2009){Bezanson}, {van Dokkum}, {Tal},
  {Marchesini}, {Kriek}, {Franx}, \& {Coppi}}]{Bezanson2009}
{Bezanson} R., {van Dokkum} P.~G., {Tal} T., {Marchesini} D., {Kriek} M.,
  {Franx} M., {Coppi} P., 2009, \apj, 697, 1290

\bibitem[{{Binggeli} {et~al}\mbox{.}(1984){Binggeli}, {Sandage}, \&
  {Tarenghi}}]{Binggeli1984}
{Binggeli} B., {Sandage} A., {Tarenghi} M., 1984, \aj, 89, 64

\bibitem[{{Boylan-Kolchin} {et~al}\mbox{.}(2006){Boylan-Kolchin}, {Ma}, \&
  {Quataert}}]{Boylan-Kolchin2006}
{Boylan-Kolchin} M., {Ma} C.-P., {Quataert} E., 2006, \mnras, 369, 1081

\bibitem[{{Bruzual} \& {Charlot}(2003)}]{bruzual03}
{Bruzual} G., {Charlot} S., 2003, \mnras, 344, 1000

\bibitem[{{Burstein} {et~al}\mbox{.}(1987){Burstein}, {Davies}, {Dressler},
  {Faber}, {Stone}, {Lynden-Bell}, {Terlevich}, \& {Wegner}}]{Burstein1987}
{Burstein} D., {Davies} R.~L., {Dressler} A., {Faber} S.~M., {Stone} R.~P.~S.,
  {Lynden-Bell} D., {Terlevich} R.~J., {Wegner} G., 1987, \apjs, 64, 601

\bibitem[{{Caon} {et~al}\mbox{.}(1993){Caon}, {Capaccioli}, \&
  {D'Onofrio}}]{Caon1993}
{Caon} N., {Capaccioli} M., {D'Onofrio} M., 1993, \mnras, 265, 1013

\bibitem[{{Cappellari} {et~al}\mbox{.}(2006){Cappellari}, {Bacon}, {Bureau},
  {Damen}, {Davies}, {de Zeeuw}, {Emsellem}, {Falc{\'o}n-Barroso},
  {Krajnovi{\'c}}, {Kuntschner}, {McDermid}, {Peletier}, {Sarzi}, {van den
  Bosch}, \& {van de Ven}}]{Cappellari2006}
{Cappellari} M. {et~al.}, 2006, \mnras, 366, 1126

\bibitem[{{Cappellari} {et~al}\mbox{.}(2009){Cappellari}, {di Serego
  Alighieri}, {Cimatti}, {Daddi}, {Renzini}, {Kurk}, {Cassata}, {Dickinson},
  {Franceschini}, {Mignoli}, {Pozzetti}, {Rodighiero}, {Rosati}, \&
  {Zamorani}}]{Cappellari2009}
{Cappellari} M. {et~al.}, 2009, \apjl, 704, L34

\bibitem[{{Cappellari} {et~al}\mbox{.}(2011{\natexlab{a}}){Cappellari},
  {Emsellem}, {Krajnovi{\'c}}, {McDermid}, {Scott}, {Verdoes Kleijn}, {Young},
  {Alatalo}, {Bacon}, {Blitz}, {Bois}, {Bournaud}, {Bureau}, {Davies}, {Davis},
  {de Zeeuw}, {Duc}, {Khochfar}, {Kuntschner}, {Lablanche}, {Morganti}, {Naab},
  {Oosterloo}, {Sarzi}, {Serra}, \& {Weijmans}}]{Cappellari2011a}
{Cappellari} M. {et~al.}, 2011{\natexlab{a}}, \mnras, 413, 813

\bibitem[{{Cappellari} {et~al}\mbox{.}(2011{\natexlab{b}}){Cappellari},
  {Emsellem}, {Krajnovi{\'c}}, {McDermid}, {Serra}, {Alatalo}, {Blitz}, {Bois},
  {Bournaud}, {Bureau}, {Davies}, {Davis}, {de Zeeuw}, {Khochfar},
  {Kuntschner}, {Lablanche}, {Morganti}, {Naab}, {Oosterloo}, {Sarzi}, {Scott},
  {Weijmans}, \& {Young}}]{Cappellari2011b}
{Cappellari} M. {et~al.}, 2011{\natexlab{b}}, \mnras, 416, 1680

\bibitem[{{Cappellari} {et~al}\mbox{.}(2012{\natexlab{a}}){Cappellari},
  {McDermid}, {Alatalo}, {Blitz}, {Bois}, {Bournaud}, {Bureau}, {Crocker},
  {Davies}, {Davis}, {de Zeeuw}, {Duc}, {Emsellem}, {Khochfar},
  {Krajnovi{\'c}}, {Kuntschner}, {Lablanche}, {Morganti}, {Naab}, {Oosterloo},
  {Sarzi}, {Scott}, {Serra}, {Weijmans}, \& {Young}}]{Cappellari2012}
{Cappellari} M. {et~al.}, 2012{\natexlab{a}}, \nat, 484, 485

\bibitem[{{Cappellari} {et~al}\mbox{.}(2012{\natexlab{b}}){Cappellari},
  {McDermid}, {Alatalo}, {Blitz}, {Bois}, {Bournaud}, {Bureau}, {Crocker},
  {Davies}, {Davis}, {de Zeeuw}, {Duc}, {Khochfar}, {Krajnovic}, {Kuntschner},
  {Morganti}, {Naab}, {Oosterloo}, {Sarzi}, {Scott}, {Serra}, {Weijmans}, \&
  {Young}}]{Cappellari2012p20}
{Cappellari} M. {et~al.}, 2012{\natexlab{b}}, ArXiv:1208.3523

\bibitem[{{Cappellari} {et~al}\mbox{.}(2012{\natexlab{c}}){Cappellari},
  {Scott}, {Alatalo}, {Blitz}, {Bois}, {Bournaud}, {Bureau}, {Crocker},
  {Davies}, {Davis}, {de Zeeuw}, {Duc}, {Khochfar}, {Krajnovic}, {Kuntschner},
  {McDermid}, {Morganti}, {Naab}, {Oosterloo}, {Sarzi}, {Serra}, {Weijmans}, \&
  {Young}}]{Cappellari2012p19}
{Cappellari} M. {et~al.}, 2012{\natexlab{c}}, ArXiv:1208.3522

\bibitem[{{Cassata} {et~al}\mbox{.}(2011){Cassata}, {Giavalisco}, {Guo},
  {Renzini}, {Ferguson}, {Koekemoer}, {Salimbeni}, {Scarlata}, {Grogin},
  {Conselice}, {Dahlen}, {Lotz}, {Dickinson}, \& {Lin}}]{Cassata2011}
{Cassata} P. {et~al.}, 2011, \apj, 743, 96

\bibitem[{{Cenarro} \& {Trujillo}(2009)}]{Cenarro2009}
{Cenarro} A.~J., {Trujillo} I., 2009, \apjl, 696, L43

\bibitem[{{Chabrier}(2003)}]{Chabrier2003}
{Chabrier} G., 2003, \pasp, 115, 763

\bibitem[{{Chen} {et~al}\mbox{.}(2010){Chen}, {C{\^o}t{\'e}}, {West}, {Peng},
  \& {Ferrarese}}]{Chen2010}
{Chen} C., {C{\^o}t{\'e}} P., {West} A.~A., {Peng} E.~W., {Ferrarese} L., 2010,
  \apjs, 191, 1

\bibitem[{{Cimatti} {et~al}\mbox{.}(2008){Cimatti}, {Cassata}, {Pozzetti},
  {Cappellari}, {Cappellari}, {Cappellari}, {Cappellari}, {Cappellari}, \& {et
  al.}}]{Cimatti2008}
{Cimatti} A. {et~al.}, 2008, \aap, 482, 21

\bibitem[{{Ciotti}(2009)}]{Ciotti2009}
{Ciotti} L., 2009, Nuovo Cimento Rivista Serie, 32, 1

\bibitem[{{Conroy} \& {van Dokkum}(2012)}]{Conroy2012models}
{Conroy} C., {van Dokkum} P., 2012, \apj, 747, 69

\bibitem[{{Daddi} {et~al}\mbox{.}(2005){Daddi}, {Renzini}, {Pirzkal},
  {Cimatti}, {Malhotra}, {Stiavelli}, {Xu}, {Pasquali}, {Rhoads}, {Brusa}, {di
  Serego Alighieri}, {Ferguson}, {Koekemoer}, {Moustakas}, {Panagia}, \&
  {Windhorst}}]{Daddi2005}
{Daddi} E. {et~al.}, 2005, \apj, 626, 680

\bibitem[{{Davies} {et~al}\mbox{.}(1983){Davies}, {Efstathiou}, {Fall},
  {Illingworth}, \& {Schechter}}]{Davies1983}
{Davies} R.~L., {Efstathiou} G., {Fall} S.~M., {Illingworth} G., {Schechter}
  P.~L., 1983, \apj, 266, 41

\bibitem[{{de Vaucouleurs}(1948)}]{deVaucouleurs1948}
{de Vaucouleurs} G., 1948, Annales d'Astrophysique, 11, 247

\bibitem[{{de Vaucouleurs} {et~al}\mbox{.}(1991){de Vaucouleurs}, {de
  Vaucouleurs}, {Corwin}, {Buta}, {Paturel}, \& {Fouque}}]{deVaucouleurs1991}
{de Vaucouleurs} G., {de Vaucouleurs} A., {Corwin}, Jr. H.~G., {Buta} R.~J.,
  {Paturel} G., {Fouque} P., 1991, Third Reference Catalogue of Bright
  Galaxies. Volume 1-3, XII, 2069 pp.~7 figs..~ Springer-Verlag Berlin
  Heidelberg New York

\bibitem[{{di Serego Alighieri} {et~al}\mbox{.}(2005){di Serego Alighieri},
  {Vernet}, {Cimatti}, {Cappellari}, {Cappellari}, {Cappellari}, {Cappellari},
  {Cappellari}, \& {et al.}}]{diSeregoAlighieri2005}
{di Serego Alighieri} S. {et~al.}, 2005, \aap, 442, 125

\bibitem[{{Emsellem} {et~al}\mbox{.}(2011){Emsellem}, {Cappellari},
  {Krajnovi{\'c}}, {Alatalo}, {Blitz}, {Bois}, {Bournaud}, {Bureau}, {Davies},
  {Davis}, {de Zeeuw}, {Khochfar}, {Kuntschner}, {Lablanche}, {McDermid},
  {Morganti}, {Naab}, {Oosterloo}, {Sarzi}, {Scott}, {Serra}, {van de Ven},
  {Weijmans}, \& {Young}}]{Emsellem2011}
{Emsellem} E. {et~al.}, 2011, \mnras, 414, 888

\bibitem[{{Faber} \& {Jackson}(1976)}]{Faber1976}
{Faber} S.~M., {Jackson} R.~E., 1976, \apj, 204, 668

\bibitem[{{Gallazzi} \& {Bell}(2009)}]{Gallazzi2009}
{Gallazzi} A., {Bell} E.~F., 2009, \apjs, 185, 253

\bibitem[{{Giavalisco} {et~al}\mbox{.}(2004){Giavalisco}, {Ferguson},
  {Koekemoer}, {Dickinson}, {Alexander}, {Bauer}, {Bergeron}, {Biagetti},
  {Brandt}, {Casertano}, {Cesarsky}, {Chatzichristou}, {Conselice},
  {Cristiani}, {Da Costa}, {Dahlen}, {de Mello}, {Eisenhardt}, {Erben}, {Fall},
  {Fassnacht}, {Fosbury}, {Fruchter}, {Gardner}, {Grogin}, {Hook},
  {Hornschemeier}, {Idzi}, {Jogee}, {Kretchmer}, {Laidler}, {Lee}, {Livio},
  {Lucas}, {Madau}, {Mobasher}, {Moustakas}, {Nonino}, {Padovani}, {Papovich},
  {Park}, {Ravindranath}, {Renzini}, {Richardson}, {Riess}, {Rosati},
  {Schirmer}, {Schreier}, {Somerville}, {Spinrad}, {Stern}, {Stiavelli},
  {Strolger}, {Urry}, {Vandame}, {Williams}, \& {Wolf}}]{Giavalisco2004}
{Giavalisco} M. {et~al.}, 2004, \apjl, 600, L93

\bibitem[{{Hernquist} {et~al}\mbox{.}(1993){Hernquist}, {Spergel}, \&
  {Heyl}}]{Hernquist1993}
{Hernquist} L., {Spergel} D.~N., {Heyl} J.~S., 1993, \apj, 416, 415

\bibitem[{{Hopkins} {et~al}\mbox{.}(2009){Hopkins}, {Bundy}, {Murray},
  {Quataert}, {Lauer}, \& {Ma}}]{Hopkins2009compact}
{Hopkins} P.~F., {Bundy} K., {Murray} N., {Quataert} E., {Lauer} T.~R., {Ma}
  C.-P., 2009, \mnras, 398, 898

\bibitem[{{Jorgensen} {et~al}\mbox{.}(1995){Jorgensen}, {Franx}, \&
  {Kjaergaard}}]{Jorgensen1995phot}
{Jorgensen} I., {Franx} M., {Kjaergaard} P., 1995, \mnras, 273, 1097

\bibitem[{{Kormendy}(1977)}]{Kormendy1977}
{Kormendy} J., 1977, \apj, 218, 333

\bibitem[{{Krajnovi{\'c}} {et~al}\mbox{.}(2011){Krajnovi{\'c}}, {Emsellem},
  {Cappellari}, {Cappellari}, {Cappellari}, {Cappellari}, {Cappellari},
  {Cappellari}, \& {et al.}}]{Krajnovic2011}
{Krajnovi{\'c}} D. {et~al.}, 2011, \mnras, 414, 2923

\bibitem[{{Kroupa}(2001)}]{Kroupa2001}
{Kroupa} P., 2001, \mnras, 322, 231

\bibitem[{{Kroupa}(2002)}]{Kroupa2002}
{Kroupa} P., 2002, Science, 295, 82

\bibitem[{{Longhetti} {et~al}\mbox{.}(2007){Longhetti}, {Saracco},
  {Severgnini}, {Della Ceca}, {Mannucci}, {Bender}, {Drory}, {Feulner}, \&
  {Hopp}}]{Longhetti2007}
{Longhetti} M. {et~al.}, 2007, \mnras, 374, 614

\bibitem[{{Maraston}(2005)}]{Maraston2005}
{Maraston} C., 2005, \mnras, 362, 799

\bibitem[{{Maraston} {et~al}\mbox{.}(2010){Maraston}, {Pforr}, {Renzini},
  {Daddi}, {Dickinson}, {Cimatti}, \& {Tonini}}]{Maraston2010}
{Maraston} C., {Pforr} J., {Renzini} A., {Daddi} E., {Dickinson} M., {Cimatti}
  A., {Tonini} C., 2010, \mnras, 407, 830

\bibitem[{{Mihos} \& {Hernquist}(1994)}]{Mihos1994burst}
{Mihos} J.~C., {Hernquist} L., 1994, \apjl, 431, L9

\bibitem[{{Naab} {et~al}\mbox{.}(2009){Naab}, {Johansson}, \&
  {Ostriker}}]{Naab2009}
{Naab} T., {Johansson} P.~H., {Ostriker} J.~P., 2009, \apjl, 699, L178

\bibitem[{{Newman} {et~al}\mbox{.}(2012){Newman}, {Ellis}, {Bundy}, \&
  {Treu}}]{Newman2012}
{Newman} A.~B., {Ellis} R.~S., {Bundy} K., {Treu} T., 2012, \apj, 746, 162

\bibitem[{{Newman} {et~al}\mbox{.}(2010){Newman}, {Ellis}, {Treu}, \&
  {Bundy}}]{Newman2010}
{Newman} A.~B., {Ellis} R.~S., {Treu} T., {Bundy} K., 2010, \apjl, 717, L103

\bibitem[{{Nipoti} {et~al}\mbox{.}(2003){Nipoti}, {Londrillo}, \&
  {Ciotti}}]{Nipoti2003}
{Nipoti} C., {Londrillo} P., {Ciotti} L., 2003, \mnras, 342, 501

\bibitem[{{Onodera} {et~al}\mbox{.}(2012){Onodera}, {Renzini}, {Carollo},
  {Cappellari}, {Mancini}, {Strazzullo}, {Daddi}, {Arimoto}, {Gobat}, {Yamada},
  {McCracken}, {Ilbert}, {Capak}, {Cimatti}, {Giavalisco}, {Koekemoer}, {Kong},
  {Lilly}, {Motohara}, {Ohta}, {Sanders}, {Scoville}, {Tamura}, \&
  {Taniguchi}}]{Onodera2012}
{Onodera} M. {et~al.}, 2012, \apj, 755, 26

\bibitem[{{Salpeter}(1955)}]{Salpeter1955}
{Salpeter} E.~E., 1955, \apj, 121, 161

\bibitem[{{Sersic}(1968)}]{Sersic1968}
{Sersic} J.~L., 1968, Atlas de galaxias australes. Cordoba, Argentina:
  Observatorio Astronomico, 1968

\bibitem[{{Springel} {et~al}\mbox{.}(2005){Springel}, {White}, {Jenkins},
  {Cappellari}, {Cappellari}, {Cappellari}, {Cappellari}, {Cappellari},
  {Cappellari}, \& {et al.}}]{Springel2005nat}
{Springel} V. {et~al.}, 2005, \nat, 435, 629

\bibitem[{{Szomoru} {et~al}\mbox{.}(2012){Szomoru}, {Franx}, \& {van
  Dokkum}}]{Szomoru2012}
{Szomoru} D., {Franx} M., {van Dokkum} P.~G., 2012, \apj, 749, 121

\bibitem[{{Toft} {et~al}\mbox{.}(2012){Toft}, {Gallazzi}, {Zirm}, {Wold},
  {Zibetti}, {Grillo}, \& {Man}}]{Toft2012}
{Toft} S., {Gallazzi} A., {Zirm} A., {Wold} M., {Zibetti} S., {Grillo} C.,
  {Man} A., 2012, \apj, 754, 3

\bibitem[{{Toft} {et~al}\mbox{.}(2007){Toft}, {van Dokkum}, {Franx}, {Labbe},
  {F{\"o}rster Schreiber}, {Wuyts}, {Webb}, {Rudnick}, {Zirm}, {Kriek}, {van
  der Werf}, {Blakeslee}, {Illingworth}, {Rix}, {Papovich}, \&
  {Moorwood}}]{Toft2007}
{Toft} S. {et~al.}, 2007, \apj, 671, 285

\bibitem[{{Trujillo} {et~al}\mbox{.}(2006){Trujillo}, {Feulner}, {Goranova},
  {Hopp}, {Longhetti}, {Saracco}, {Bender}, {Braito}, {Della Ceca}, {Drory},
  {Mannucci}, \& {Severgnini}}]{Trujillo2006}
{Trujillo} I. {et~al.}, 2006, \mnras, 373, L36

\bibitem[{{van de Sande} {et~al}\mbox{.}(2011){van de Sande}, {Kriek}, {Franx},
  {van Dokkum}, {Bezanson}, {Whitaker}, {Brammer}, {Labb{\'e}}, {Groot}, \&
  {Kaper}}]{vandeSande2011}
{van de Sande} J. {et~al.}, 2011, \apjl, 736, L9+

\bibitem[{{van der Marel} \& {van Dokkum}(2007)}]{vanderMarel2007}
{van der Marel} R.~P., {van Dokkum} P.~G., 2007, \apj, 668, 756

\bibitem[{{van der Wel} \& {van der Marel}(2008)}]{vanderWel2008vdM}
{van der Wel} A., {van der Marel} R.~P., 2008, \apj, 684, 260

\bibitem[{{van Dokkum} \& {Conroy}(2010)}]{vanDokkum2010}
{van Dokkum} P.~G., {Conroy} C., 2010, \nat, 468, 940

\bibitem[{{van Dokkum} {et~al}\mbox{.}(2008){van Dokkum}, {Franx}, {Kriek},
  {Holden}, {Illingworth}, {Magee}, {Bouwens}, {Marchesini}, {Quadri},
  {Rudnick}, {Taylor}, \& {Toft}}]{vanDokkum2008}
{van Dokkum} P.~G. {et~al.}, 2008, \apjl, 677, L5

\bibitem[{{van Dokkum} {et~al}\mbox{.}(2009){van Dokkum}, {Kriek}, \&
  {Franx}}]{vanDokkum2009nat}
{van Dokkum} P.~G., {Kriek} M., {Franx} M., 2009, \nat, 460, 717

\bibitem[{{Vazdekis} {et~al}\mbox{.}(2010){Vazdekis},
  {S{\'a}nchez-Bl{\'a}zquez}, {Falc{\'o}n-Barroso}, {Cenarro}, {Beasley},
  {Cardiel}, {Gorgas}, \& {Peletier}}]{Vazdekis2010}
{Vazdekis} A., {S{\'a}nchez-Bl{\'a}zquez} P., {Falc{\'o}n-Barroso} J.,
  {Cenarro} A.~J., {Beasley} M.~A., {Cardiel} N., {Gorgas} J., {Peletier}
  R.~F., 2010, \mnras, 404, 1639

\end{thebibliography}
\end{document}